\documentclass[acmsmall]{acmart}

\pdfoutput=1
\usepackage{etex}

\usepackage{booktabs}

\usepackage[utf8]{inputenc}

\usepackage{mathptmx}
\usepackage[english]{babel}
\usepackage{multirow}
\usepackage{rotating}
\usepackage{color}
\usepackage{numprint}
\usepackage{xspace}
\usepackage{tabularx}
\usepackage{balance}
\usepackage{tikz}
\usepackage{pgfplots}
\usepgfplotslibrary{units}
\usepackage{caption}
\usepackage{framed}
\usepackage{mdframed}
\usepackage{latexsym}
\usepackage{amssymb}
\usepackage{amsmath}
\usepackage{comment}
\usepackage{subcaption}
\usepackage[inline]{enumitem}
\usepackage{xcolor}
\usepackage{algorithm}
\usepackage{algorithmic}

\usepackage{listings} 
\usepackage{textcomp}
\usepackage{cprotect}
\usepackage{tikz}
\usepackage{booktabs}
\usepackage{numprint}

\usepackage{listings}
\lstdefinelanguage{diff}{
  language=java,
  basicstyle=\ttfamily\scriptsize,
  sensitive=true,
  morecomment=[f][\color{gray}][0]{diff},
  morecomment=[f][\color{gray}][0]{index},
  morecomment=[f][\color{blue}][0]{@@},
  morecomment=[f][\color{magenta}][0]{***},
  morecomment=[f][\color{violet}][0]{!},
  morecomment=[f][\color{red!60!black}][0]{-},
  morecomment=[f][\color{green!60!black}][0]{+},
  morecomment=[f][\color{magenta}][0]{---},
  morecomment=[f][\color{magenta}][0]{+++},
  morecomment=[f][\color{gray}][0]{Binary},
  morecomment=[f][\color{gray}][0]{Only},
  morecomment=[f][\color{gray}][0]{old},
  morecomment=[f][\color{gray}][0]{new},
  morecomment=[f][\color{gray}][0]{rename},
  morecomment=[f][\color{gray}][0]{similarity},
  morecomment=[f][\color{gray}][0]{deleted},
  morecomment=[f][\color{magenta}][0]{***************},
  morecomment=[f][\color{red!60!black}][0]<,
  morecomment=[f][\color{green!60!black}][0]>,
  morecomment=[f][\color{blue}][0]{0},
  morecomment=[f][\color{blue}][0]{1},
  morecomment=[f][\color{blue}][0]{2},
  morecomment=[f][\color{blue}][0]{3},
  morecomment=[f][\color{blue}][0]{4},
  morecomment=[f][\color{blue}][0]{5},
  morecomment=[f][\color{blue}][0]{6},
  morecomment=[f][\color{blue}][0]{7},
  morecomment=[f][\color{blue}][0]{8},
  morecomment=[f][\color{blue}][0]{9},
}[comments]

\usepackage[]{pdfcomment}

\usetikzlibrary{shapes, arrows}

\tikzset{
  cross/.style={path picture={\draw[red]
        (path picture bounding box.south east) -- (path picture bounding box.north west)
        (path picture bounding box.south west) -- (path picture bounding box.north east);}}
}

\newcommand{\failing}[1][red,fill=white]{\tikz[baseline=-0.7ex]\draw[#1,radius=3pt,cross] (0,0) circle ;\xspace}
\newcommand{\passing}[1][green!60!black,fill=green!60!black]{\tikz[baseline=-0.7ex]\draw[#1,radius=3pt] (0,0) circle ;\xspace}

\newcommand{\validPatch}{
}

\newcommand{\TODO}[1]{{\color{red}{TODO: #1}}\GenericWarning{}{LaTeX Warning: TODO: #1}}\newcommand\todo\TODO
\newcommand{\question}[2]{{\bf RQ#1. #2}}
\newcommand{\answer}[2]{\vspace{.3cm}{\centering\fbox{\parbox{0.99\columnwidth}{\textbf{Answer to RQ#1}. #2}}}\vspace{.3cm}}

\AtBeginDocument{

}

\newcommand\generationNbBug{34\xspace}

\newcommand\itzal{{Itzal}\xspace}
\newcommand\tool{Itzal4j\xspace}

\newcommand\productionApplication{Unmodified Application\xspace}
\newcommand\shadowService{Shadower\xspace}
\newcommand\patchService{Patch Generation Service\xspace}
\newcommand\regressionService{Regression Assessment Service\xspace}

\newcommand\canditatePatches{{candidate patches}\xspace} 
\newcommand\canditatePatch{{candidate patch}\xspace}

\newcommand\requestOracle{{Request-oracle}\xspace}

\acmJournal{PACMPL}
\acmVolume{1}
\acmNumber{OOPSLA} 
\acmArticle{1}
\acmYear{2018}
\acmMonth{1}
\acmDOI{} 
\startPage{1}

\setcopyright{none}

\begin{document}
\title{Production-Driven Patch Generation and Validation}

\author{Thomas Durieux}
\affiliation{
  \institution{Inria \& University of Lille}
  \city{Lille}
  \state{France}
}

\author{Youssef Hamadi}
\affiliation{
  \institution{Ecole Polytechnique}
  \city{Paris}
  \state{France}
}

\author{Martin Monperrus}
\affiliation{
  \institution{KTH Royal Institute of Technology}
  \city{Stockholm}
  \state{Sweden}
}

\begin{abstract}
We envision a world where the developer would receive each morning in her GitHub dashboard a list of potential patches that fix certain production failures.
For this, we propose a novel program repair scheme, with the unique feature of being applicable to production directly.
We present the design and implementation of a prototype system for Java, called \itzal, that performs patch generation for uncaught exceptions in production.
We have performed two empirical experiments to validate our system: the first one on \generationNbBug failures from 14 different software applications, the second one on 16 seeded failures in 3 real open-source e-commerce applications for which we have set up a realistic user traffic. 
This validates the novel and disruptive idea of using program repair directly in production.
\end{abstract}

\maketitle

\section{Introduction}
In modern distributed systems running on the cloud, software failures happen constantly \cite{oppenheimer2003internet}. 
The leading company in the business of production failure monitoring, called OverOps, has reported that a popular Java web application suffers from 9.2 million exceptions per month on average, due to an average of 53 unique root causes \cite{takipi}.

What about automatically generating source-code patches that would prevent production failures from happening again? We dream of a world where the developer would receive each morning in her GitHub dashboard a list of potential patches that fix certain production failures.
This is the blue-sky vision we elaborate in this paper.

This is fundamentally different from traditional program repair (e.g. \cite{le2012genprog,semfix}). Indeed, traditional program repair is built on premises that are not adequate to fix production failures.
First, most repair systems require one or several failing test cases to guide the repair process. But it has been shown that it is extremely difficult to reproduce production failures and translate them into failing test cases \cite{reproducedifficulty1,reproducedifficulty2,reproducedifficulty3}. 
Second, traditional program repair uses a regression test suite to verify that the generated patch has not introduced regressions, with no guarantee whatsoever that the regression test suite covers all the behaviors used in production, resulting in incorrect patches \cite{kali,Smith15fse,martinez2016automatic}.

There is a fundamental gap between the vision of automatically generating patches for production failures and the state-of-the-art of program repair. This is what we address here -- we bring program repair to production failures. 
In this paper, we specify a novel scheme for program repair in production, and we present the design and implementation of a prototype system for Java, called \itzal. 

\itzal works as follows.
First, it uses production oracles (such as uncaught exceptions) to detect failures and trigger patch search.
Second, right after the failure is detected in production, a patch is searched in a parallel environment that mimics the production one. This search is asynchronous so that patch synthesis has a negligible overhead on the production system.
Third, the synthesized patches are validated based on traffic that is an exact copy of the production user traffic -- we call it shadow traffic.
Patches that fix a production failure and do not introduce regressions that are visible on the end-to-end user traffic are proposed to the developer. 

To show that the \itzal vision is feasible, we have implemented it for Java.
To demonstrate that it is generic, we have considered two different patch models:
one for null pointer exceptions (\#1 exception in production \cite{takipi}), and one for arbitrary uncaught runtime exceptions.
We have performed two large scale experiments: the first one on \generationNbBug failures from 14 different software applications used in production, the second one on 16 seeded failures in 3 real open-source e-commerce applications for which we have set up a realistic user traffic. Those experiments validate our novel concept of patch synthesis in production.

In summary, our contributions are:
\begin{itemize}
\item A novel scheme for program repair in production, it performs patch synthesis and regression testing live based on user traffic. This is a fundamental breakthrough with traditional program repair based on tests or static analysis that may open a new research avenue.

\item The design and implementation of a prototype in Java, called \itzal. It is made open-source for fostering follow-up research in that new area.

\item The evaluation over \generationNbBug production failures from 14 different large scale software applications. It is the first
proof-of-concept that our vision of program repair in production is feasible.

\end{itemize}

The remainder of this paper is organized as follows.
\autoref{sec:challenges}  presents the current challenges of patch generation.
\autoref{sec:contribution} presents \itzal.
\autoref{sec:evaluation} presents the evaluation of \itzal.
\autoref{sec:rw} presents the related works and \autoref{sec:conclusion} concludes.

\section{Challenges of Automatic Software Repair}
\label{sec:challenges}

Automatic software repair is the process of fixing software bugs automatically. 
However, this broad definition covers techniques that are fundamentally different.
First, ``repair'' means patching the program code, and the literature \cite{monperrus2015} mainly refers to this as ``program repair'', ``patch generation'' or colloquially ``automatic bug fixing''. An archetypal program repair system is GenProg \cite{le2012genprog}.
Second, repair also means changing the program state in response to field failures at runtime. The literature uses ``runtime repair'', ``self-healing'' and ``failure-oblivious computing'' to refer to this idea. An archetypal self-healing system is ClearView \cite{perkins2009automatically}. 

Note that there is a key difference here: runtime repair does not mean generating source code patch as done in patch generation.
Runtime repair in production means modifying the execution.
For example, ClearView does not generate source code patches, it restores CPU register invariants at runtime. 
Reversely, GenProg cannot be used in production, since it takes as input test cases.

\subsection{Challenges of Patch Generation}

In production, end-users trigger bugs by exercising the system with inputs and sequences of interactions that were unforeseen by the development team. 
When this happens, they may report the bug to the development team through a support channel or a bug tracking system. 
The problem is that there is a huge gap between an issue reported by an end-user and a failing test case that is usable by a test-suite based repair system. The main problem is that the issue is reported in natural language by end users, and often the description is incomplete. 
The extreme difficulty of reproducing production bugs has been confirmed by numerous studies (e.g., \cite{reproducedifficulty1,reproducedifficulty2,reproducedifficulty3}) and little support is available today for this task. Thus, current patch generation systems are inapplicable for production failures, because much manual work is needed to translate an end-user problem into a valid failing test case. Similarly, it is also difficult to convert production traffic into regression tests.

In this paper, we recast patch generation in a unique and novel way: we bring patch generation to production.
\subsection{Challenge of Self-Healing Systems}

Self-healing software systems have completely different challenges.
The core idea of self-healing software is to modify the execution in response to a runtime failure.
The first and foremost challenge of self-healing systems is to assess whether the state after healing is viable. Since self-healing software replaces a crashing failure into a continued execution, this execution is speculative in essence and no programmer has designed it upfront. There is very little work on assessing the viability of program states after self-healing \cite{locasto2008life}.
The second challenge is to transform those runtime modifications into valuable knowledge for developers. One solution is to log every execution modification, but this may represent an enormous volume of logs for any production system with large traffic, and this volume would be of little use for developers.

In this paper, the novelty we propose is that developers receive source code patches directly synthesized from production executions.

\section{Program Repair on Production Failures}\label{sec:contribution}

\begin{figure}[t!]
\centering
\includegraphics[width=0.90\columnwidth]{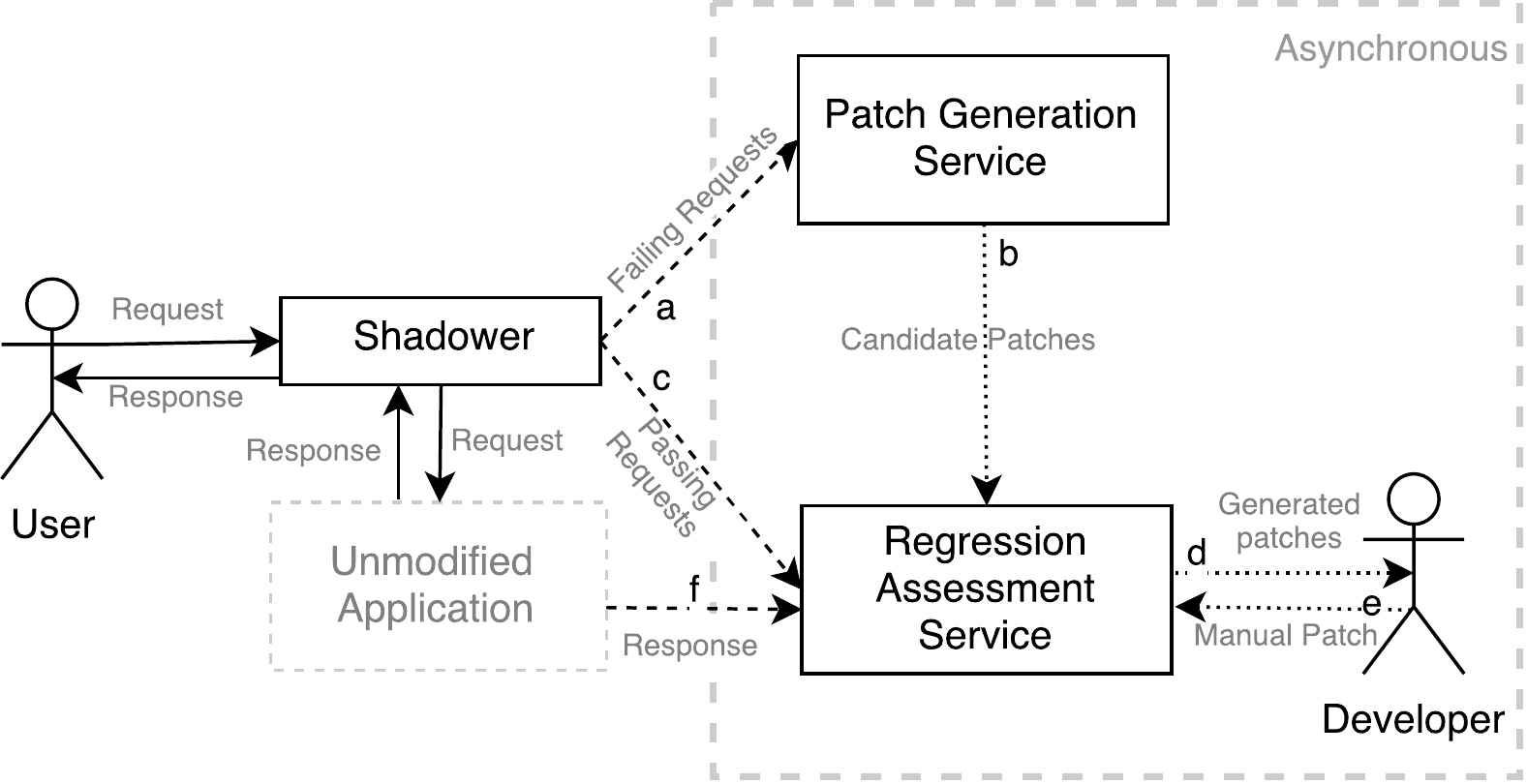} 
\caption{The blueprint of \itzal. The key idea is to duplicate user traffic via a ``shadower'', the duplicated traffic is used to search for patches and to validate candidate patches.}

\label{fig:architecture}
\end{figure}

We now present \itzal, a novel program repair technique for generating patches directly in the production environment.

\emph{Intuition: }
The intuition behind \itzal is twofold. First, one can use production runtime contracts to drive the generation of source code patches. This includes classical pre- and post-conditions as well as implicit contracts such as that an accessed variable must not be null. The latter is important because the violations of those implicit contracts come for free in any modern runtime, usually in the form of runtime exceptions. The second intuition is that one can use the diversity of the production inputs to perform in-the-field regression testing on the synthesized patches. 

\emph{Applicability: }
The requirement to deploy \itzal is that the application has a message-driven architecture \cite{reactivemanifesto}, i.e., must use requests.
The type of requests may vary between applications, it can be for example:
1) a request sent by a user's browser to a webserver,
2) a REST message for a micro-service application,
3) for a mobile application, a touch event triggered when a user touches a mobile device's screen.
An extreme case of message-driven software is serverless computing, also known as Function-as-a-Service \cite{adzic2017serverless}, such as Amazon Lambda, where there is no state between requests. One may consider that Function-as-a-Service is a killer application domain of \itzal. 

\subsection{Blueprint Architecture}

The \itzal architecture is composed of three main components that are set up around an existing unmodified production application, as shown in \autoref{fig:architecture}.
\begin{enumerate}[leftmargin=*]

\item The \shadowService (see \autoref{sec:shadow}) is used to duplicate the requests of the \productionApplication. The duplicated requests are then sent in parallel to the \patchService and the \regressionService.

\item The \patchService (see \autoref{sec:patch_generation}) is the service that searches for patches that fix a given failure. It uses a failure detection mechanism called ``\requestOracle'' in this paper to determine whether the application has successfully handled a request.

\item The \regressionService (see \autoref{sec:patch_regression}) performs regression testing on the patches based on user traffic. It applies the generated patches on a copy of the application  -- the shadow application -- and \shadowService duplicates the user traffic to the shadow application in order to observe the behavior of the patched application and potentially detect regressions.
\end{enumerate}

Eventually, the patches generated by \itzal are communicated to the developers, for instance using automated pull requests on GitHub. The developers can directly merge them or further improve them.

\emph{Algorithm: } \autoref{algo:main} shows the workflow of \itzal.
\shadowService receives the request from the client (line \autoref{algo:itzal:request}).
Then it redirects the request to the \productionApplication (line \autoref{algo:itzal:send_application}).
Once the request has been handled by the \productionApplication, 
the response is sent back to the client (line 3). 
If the \requestOracle has determined that there is a failure, the request is sent to the \patchService (arrow $a$ in \autoref{fig:architecture} and line \autoref{algo:itzal:send_patch}).
The patches generated by \patchService that pass the \requestOracle (i.e., that fix the failure at hand) are sent to \regressionService (line \autoref{algo:itzal:push_patch}).
If the request has succeeded (i.e., no failure on the original application), the request is also sent to the \regressionService (line \autoref{algo:itzal:send_regression}) where all the previously generated patches are being regressed on-the-fly against the new requests.
When the \regressionService has identified patches with no regressions, it sends them to the developers.

\begin{algorithm}[t]
  \begin{algorithmic}[1]
    \REQUIRE{A: the \productionApplication}
    \REQUIRE{G: the \patchService}
    \REQUIRE{V: the \regressionService}
    \WHILE{new request $r$ from Client}\label{algo:itzal:request}
        \STATE{$output$ $\leftarrow$ A($r$)} \label{algo:itzal:send_application}
        \STATE{send $output$ to Client}
        \IF{$r$ produces a failure}\label{algo:itzal:oracle}
            \STATE{$patches$ $\leftarrow$ G($r$)} \label{algo:itzal:send_patch}
            \STATE{$failureCount_r++$} 
            \STATE{push $patches$ to V} \label{algo:itzal:push_patch}
        \ELSE
            \STATE{send ($r$,$output$) to V for regression analysis} \label{algo:itzal:send_regression}
        \ENDIF
        \IF{$\exists$ not ``regressive patches'' $\in V$}\label{algo:validation}
             \STATE{report patches to developers} \label{algo:report}
        \ENDIF
    \ENDWHILE
  \end{algorithmic}
  \caption{The main \itzal algorithm}
  \label{algo:main}
\end{algorithm}

\subsubsection{\patchService} \label{sec:patch_generation}

For every request, \itzal verifies whether the application has succeeded to answer the request using a \requestOracle.
For instance, in a webserver, one can check the return code of HTTP request (``assert response\_code != 5XX (internal server error)'') or check the presence or not of an exception.
\itzal works with generic oracles such as checking the absence of exceptions (e.g., in a web request container or in a thread monitor), and it can also work with domain-specific oracles written by software engineers on top of domain concepts and data (e.g., the returned XML must comply with a specific schema).

For each failing request, 
a \patchService searches for patches that prevent the failure or any other ones from happening according to a patch model.
On this, \itzal piggy-backs on existing research \cite{rinard2004enhancing,npefix}.
\itzal is agnostic to the patch service, it  supports several patch models: we have implemented two of them, both used in our evaluation.

\textbf{Definition: } a ``patch model'' enables one to enumerate all patches according to a specification of the search space. 

For each explored candidate patch, the \patchService calls \requestOracle to verify that the request has been correctly handled by the patch under consideration, i.e., the failure has been fixed.
As the \patchService generates the patches based on only one request (the failing one), the patches may break the behavior of the application for other requests, i.e., they may introduce regressions.
Thus, if the patch is successful on the failing request, it is transferred to the \regressionService (arrow $b$ in \autoref{fig:architecture}) which will further regress it on incoming requests.

The execution of candidate patches can change the state of the application in runtime.
To nullify the potential side effects of the request or the new behavior introduced in synthesized patches, each execution is done in a completely sandboxed environment.
In other words, the side-effects of the execution of the patch candidate never propagate to the production application by construction.

Beyond null dereferences, \itzal can work with any patch model, whether domain-specific (e.g., for out-of-bounds exception \cite{pldioutofbound}) or generic (e.g., Genprog \cite{le2012genprog}). Note if the patch model generates too many patches (i.e., the search space is too large), it can possibly be a problem as it can incur a huge computation overhead on the \patchService and much more importantly on the \regressionService.

\subsubsection{\regressionService} \label{sec:patch_regression}

The patches generated by the \patchService can introduce regressions as their generation involves only the failing request.
The \regressionService has the responsibility to check the behaviors of the application when the generated patches are injected on other requests.

The \regressionService uses an ``execution comparison oracle'' to compare the output of the \productionApplication with that of a patched version for the same request.
If the outputs are different, the \regressionService discards the patch and marks it as a ``regressive patch''.
For example, an execution comparison oracle for a web server can compare the HTML texts of both versions.

\textbf{Definition: } an ``execution comparison oracle'' is a mechanism to detect changed behaviors in production.

The comparison is not necessarily a byte-to-byte one, it can include heuristics to discard transient information such as time, cookie identifiers, etc.
To increase the accuracy of the regression evaluation, each generated patch is evaluated against a large number of requests, say for example 1 million if there are a large number of users.

The comparison is made on-the-fly, directly on user traffic. Doing regression testing ``live'' has the advantage that there is no need to record the potentially enormous amount of production data.

\autoref{algo:regression} is the main algorithm of the \regressionService.
The \regressionService requires a copy of the \productionApplication, the response of the \productionApplication, an execution comparison oracle, and a list of patches to regress (sent previously by the \patchService).

For each successful request received from the \shadowService (arrow $c$ in \autoref{fig:architecture} and line \autoref{algo:regression:new_request} in \autoref{algo:regression}), the \regressionService iterates over each patch to detect regressions (lines 3-12) in \autoref{algo:regression}).
Finally, the patches that pass user traffic based regression testing are sent to the developers (arrow $d$ in \autoref{fig:architecture} and line 13 in \autoref{algo:regression}).

There is a major advantage of doing regression validation on user traffic:
the user traffic contains far more usage scenarios and far more diverse values than a regression test suite.
Consequently, it reduces the risk of overfitting, i.e. it reduces the risk of suggesting an incorrect patch to the developer.

We note that the  \regressionService can also be used to validate a human patch with the production traffic, as shown in arrow $e$ of \autoref{fig:architecture}. In this case, it means that the \regressionService is used for live testing of code on user traffic.

\begin{algorithm}[t]
  \caption{The \regressionService algorithm}
  \label{algo:regression}
  \begin{algorithmic}[1]
    \REQUIRE{A: \productionApplication}
    \REQUIRE{R: Execution Comparison Oracle}
    \REQUIRE{Q: patches from \patchService}
    \WHILE{new request $r$ from Shadower}\label{algo:regression:new_request}
             \STATE{$output_{ref} \leftarrow$ the output of A for $r$ (from \shadowService)}
             \FOR{patch $p$ in Q} \label{algo:regression:patch_iteration}
                \STATE{$A' \leftarrow$ apply $p$ to A}
                \STATE{$(output_{A'})$ $\leftarrow$ $r$ send to $A'$}

                \STATE{S $\leftarrow$ R($output_{ref}$, $output_{A'}$)}
                \IF{S = false}
                    \STATE{remove $p$ from Q}
                \ELSE
                    \STATE{$regressionSuccessCount_p++$} 
                \ENDIF
            \ENDFOR
            \STATE{send \{Q, $regressionSuccessCount_p$\} to the developer}
    \ENDWHILE
  \end{algorithmic}
\end{algorithm}

\subsubsection{\shadowService} \label{sec:shadow}

The role of the \shadowService is to create shadow traffic from actual end-user traffic coming into the application.
The ``shadow traffic'' is made up of production requests that are duplicated one or several times and sent to sandboxed shadow applications.
In our case, the shadow applications are the \patchService and \regressionService.

In \itzal, the \shadowService receives the requests from the clients, duplicates them and sends one duplicate to each service of the architecture  (arrows $a$, $c$ in \autoref{fig:architecture}). The response is also shadowed for the \regressionService (arrow $f$ in \autoref{fig:architecture}).

\textbf{Definition: } a ``shadower'' is a system to duplicate requests of message-driven application.

\textbf{Definition: } a ``shadow application'' is a duplicate and sandboxed copy of a production application, which receives the same requests.

If the production application has a state (typically stored in a database), the shadow application accesses to the production data through a read-only database connection\footnote{this is supported by all major databases, whether relational or NoSQL}.
This guarantees that the shadow application never corrupts the production state, and that patch synthesis remains transparent and safe for the unmodified, deployed application.
The drawback is that it prevents repair of code related to state modification.
There are sophisticated ways for overcoming this limitation, but this is a hard and unresearched problem which is left to future work.

In the context of web applications, the concept of running multiple instances of an application is well known and heavily used: this is done for load balancing and rolling deployment. 
The difference between a load balancer and a \shadowService is twofold:
first, a load balancer does not duplicate the traffic;
second, a load balancer does not send requests to sandboxed ``sinks'' as \itzal does.

Since \itzal is a production technique, it must have a reasonable impact on the performance of the application.
In order to minimize the impact on the \productionApplication, \itzal computes the \regressionService and the \patchService asynchronously.
Indeed, the goal of \itzal is to perform patch generation, it is not an automatic error recovery system.
Hence, the \shadowService directly sends the output as soon as the  \productionApplication has handled a request (even if there is a failure).
\itzal does not have to wait for the end of the patch search or the regression testing for sending the response back to the client.
Consequently, the \shadowService is the only component that impacts the performance of the \productionApplication. 

In a typical HTTP-based setup, the cost of copying and rerouting requests on-the-fly is similar to that of classical web proxies and load balancers, which are extensively used in production systems.

\subsection{Prototype Implementation for Java}

We have implemented a prototype of \itzal for Java in a tool named \tool, which is dedicated to message-driven applications based on HTTP.
For sake of open-science, \tool is publicly available on \url{http://anonymous.4open.science/repository/f5722a25-5510-4dd7-bd04-664d5f47715a/}.

\tool has a default \requestOracle based on unhandled exceptions.
Any unhandled exception happening during the processing of a request is considered as a failure.
Also, for the case studies in the domain of web applications that will be presented later in \autoref{sec:evaluation}, we have also implemented a \requestOracle based on HTTP return codes.
According to the specification of the HTTP status codes, the HTTP status code that begins with the digit ``5'' indicates that the server is aware that it has encountered an error.
Failure detection is achieved by checking whether the HTTP status code begins with the digit ``5''. If it is the case, the request is considered as failing. Otherwise, it is considered as succeed.

\subsubsection{Implementation of the \patchService}
\label{sec:implem:path-search}

In our implementation, the \patchService uses two different patch enumeration techniques.
First, our prototype system uses the NPEFix model \cite{npefix} which addresses null dereferences in Java.
Second, our prototype system also implements the exception-stop model \cite{dobolyi2008changing}, which prevents the failure from happening by adding try-catch blocks at the method level.

\subsubsection{Implementation of the \regressionService}
The \regressionService first receives and stores a list of patches from the \patchService.
Then, it will apply the requests that it receives from the \shadowService on each patch.
Finally, the observable behavior of the patched application is compared with that of the \productionApplication, and the results of the comparison are stored to provide statistics to the developers.

In \tool, the HTTP body of the response of the \productionApplication is compared against that of the patched application (e.g., the HTML body text for web apps). 
The comparison discards transient information (e.g., IP addresses and dates).
It can further be configured with domain-specific heuristics.
If the outputs do not match, the patch is discarded and is permanently marked as a ``regressive patch''.

\subsubsection{Implementation of the \shadowService}
The \shadowService is implemented with the Jetty Proxy Library.\footnote{Jetty Proxy Servlet \url{http://www.eclipse.org/jetty/documentation/9.4.x/proxy-servlet.html}}
The major implementation challenge is to maintain a list of session identifiers (e.g. cookies) for each shadowed service. 
To achieve this, when a session-enabled request arrives with the session ID of the end user's browser, the \shadowService translates on-the-fly the session ID to each of the shadowed services (and vice-versa for the response).

\subsubsection{Implementation of the Sanboxing}
Sandboxing is achieved using the Docker container system, a major software containerization platform which provides powerful sandboxing capability (including both disk and network sandboxing) \cite{dockersandboxing}. The \patchService and the \regressionService are encapsulated in their respective docker images, with disk and network sandboxing enabled, so that it is impossible for them to impact the production state.

\subsubsection{Communication with the Developer}
\label{sec:dashboard}

Now we discuss the communication of the patches with the developers (arrow $d, e$ in \autoref{fig:architecture}).
In the current prototype, we use a web dashboard where the developers follow in real time the failures, the generated patches and the progression of the regression testing of the patches on user traffic. 
We also imagine an approach integrated into the versioning system (Git/GitHub) where patches are communicated to the developers with automated pull requests.

If patches exist for multiple failures, we use the failure count $failureCount_r$ from \autoref{algo:main} to order the patches.
The idea is that the developer would prefer to spend time firstly to the most frequent failures.
It also happens that, for the same failure kind, multiple patches successfully pass the regression testing done over the user traffic.
Consequently, we also sort the patches in order to first propose the most useful ones to the developers.
We prioritize the patches according to the regression success count ($regressionSuccessCount_p$ from \autoref{algo:regression}). 
The idea is that the more a patch has been executed by the \regressionService, the more confidence we have in it.

\section{Evaluation}\label{sec:evaluation} 

In this section, we demonstrate the feasibility of our novel and disrupting scheme of patch synthesis in production.
Our blueprint architecture addresses many different aspects:
patch generation, regression detection based on shadow traffic, and shadower.

\begin{figure}
\includegraphics[width=0.80\columnwidth]{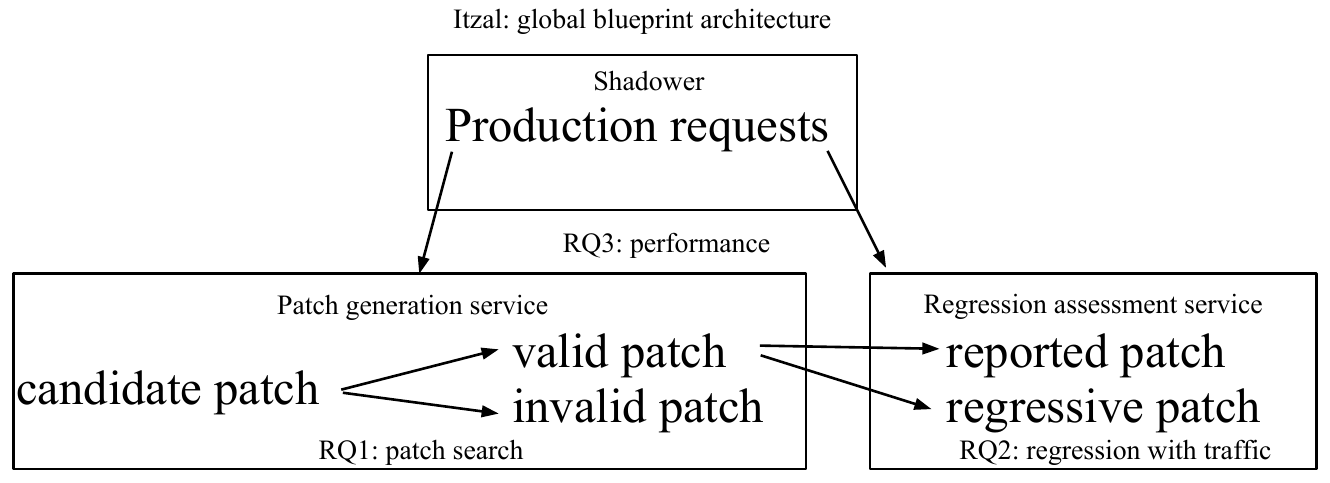}
\caption{Our research questions target each component in isolation as well as the global end-to-end approach.
}
\label{fig:rqs}
\end{figure}

We devise a research protocol that aims at:
1) studying each aspect one by one in isolation,
and 2) studying the \itzal{} from an end-to-end perspective. \autoref{fig:rqs} depicts the evaluation approach.

\newcommand\rqGeneration{\question{1}{[Live Patch Synthesis Feasibility] To what extent is it possible to generate patches in production, directly from failing requests triggered by user traffic?}\xspace}
\rqGeneration
The research question aims at evaluating the \patchService to verify that it is possible to generate patches directly from failures in a production environment, where failing requests replace failing test cases.

\newcommand\rqRegression{\question{2}{[User Traffic Effectiveness for Regression] What is the effectiveness of using user-traffic to perform live regression testing?}\xspace}
\rqRegression
The research question aims to evaluate the \regressionService. We want to see whether one can use user traffic to discard incorrect patches.
We study the effectiveness of four execution comparison oracles which are used to compare the behaviors of the patched application against that of the original application.

\newcommand\rqOverhead{\question{3}{[Performance] What is the performance overhead of \itzal?}\xspace}
\rqOverhead
The research question aims to evaluate the performance overhead of \itzal.
We measure the performance overhead of our blueprint architecture on a production system, and compute the required time needed by the \patchService to generate the patches.

\newcommand\rqEndToEnd{\question{4}{[End-to-end Effectiveness] How does \itzal work in a production-like setting?}\xspace}
\rqEndToEnd
While RQ1, RQ2 and RQ3 specifically concentrate on \patchService, \regressionService and \shadowService, the research question aims to evaluate \itzal from an end-to-end perspective by considering two real bugs that are reproducible in a live server environment with emulated traffic.

\textbf{Benchmarks}: Note that the subjects required to answer each of the 4 research questions are not identical.
They all share the characteristics of being really hard to obtain. For instance, it is really difficult to collect and reproduce production failures in the laboratory.
Consequently, we build one specific evaluation benchmark per research question.
However, we have managed to have one common case across all research questions: Mayocat-231 is used as a real bug in RQ1, as a regression subject for RQ2, in the overhead measurement of RQ3, and in the end-to-end evaluation of RQ4.

\subsection{RQ1. Live Patch Generation Feasibility}
\label{sec:rq1}
\begin{table*}
  \centering
  \caption{The benchmark used in our experiments}
  \label{tab:bench-generation}
  \begin{tabularx}{0.9\textwidth}{|X|l|l|l|r|}
    \hline
    Bug & Origin  & Request Oracle & \rotatebox{90}{Bug Type}  & LOC \\
    \hline
    BroadleafCommerce 1282 & \cite{DurieuxHM16} & HTTP status & NPE & \numprint{161428} \\
    Collection 360 & \cite{npefix} & Exception & NPE & \numprint{21650} \\
    DataflowJavaSDK c06125d & \cite{long2017automatic} & Exception & NPE & \numprint{50655} \\
    Felix 4960 & \cite{npefix} & Exception & NPE & \numprint{33057} \\
    Javapoet 70b38e5 & \cite{long2017automatic} & Exception & NPE & \numprint{3884} \\
    Jetty 335500 & \cite{gu2016automatic} & HTTP status & NPE & \numprint{153789} \\
    Jongo f46f658 & \cite{long2017automatic} & Exception & NPE & \numprint{7384} \\
    Lang 20 & \cite{JustJE2014} & Exception & NPE & \numprint{49637} \\
    Lang 304 & \cite{npefix} & Exception & NPE & \numprint{17277} \\
    Lang 33 & \cite{JustJE2014} & Exception & NPE & \numprint{45444} \\
    Lang 39 & \cite{JustJE2014} & Exception & NPE & \numprint{45143} \\
    Lang 587 & \cite{npefix} & Exception & NPE & \numprint{17319} \\
    Lang 703 & \cite{npefix} & Exception & NPE & \numprint{19047} \\
    Math 1115 & \cite{npefix} & Exception & NPE & \numprint{90782} \\
    Math 1117 & \cite{npefix} & Exception & NPE & \numprint{90794} \\
    Math 290 & \cite{npefix} & Exception & NPE & \numprint{38728} \\
    Math 305 & \cite{npefix} & Exception & NPE & \numprint{38893} \\
    Math 369 & \cite{npefix} & Exception & NPE & \numprint{41082} \\
    Math 4 & \cite{JustJE2014} & Exception & NPE & \numprint{164667} \\
    Math 70 & \cite{JustJE2014} & Exception & NPE & \numprint{83720} \\
    Math 79 & \cite{JustJE2014} & Exception & NPE & \numprint{89611} \\
    Math 988A & \cite{npefix} & Exception & NPE & \numprint{82442} \\
    Math 988B & \cite{npefix} & Exception & NPE & \numprint{82443} \\
    Mayocat 231 & \cite{DurieuxHM16} & HTTP status & NPE & 31231\\
    PDFBox 2812 & \cite{npefix} & Exception & NPE & \numprint{67294} \\
    PDFBox 2965 & \cite{npefix} & Exception & NPE & \numprint{64375} \\
    PDFBox 2995 & \cite{npefix} & Exception & NPE & \numprint{64821} \\
    Sling 4982 & \cite{npefix} & Exception & NPE & 583 \\
    Tomcat 43758 & \cite{gu2016automatic} & HTTP status & NPE & \numprint{156480} \\
    Tomcat 54703 & \cite{gu2016automatic} & HTTP status & NPE & \numprint{186301} \\
    Tomcat 55454 & \cite{gu2016automatic} & HTTP status & NPE & \numprint{193648} \\
    Tomcat 56010 & \cite{gu2016automatic} & HTTP status & IAE & \numprint{195130} \\
    Tomcat 58232 & \cite{gu2016automatic} & HTTP status & NPE & \numprint{224194} \\
    Webmagic ff2f588 & \cite{long2017automatic} & Exception & NPE & \numprint{9239} \\
    \hline
    \multicolumn{4}{|l|}{34 bugs from 14 software applications} & \numprint{2622172} \\
    \hline
  \end{tabularx}
\end{table*}

\begin{table*}
  \centering
  \caption{The feasibility of using two patch generation models for production failures. Many patches from the patch models' search space are marked as invalid because they fail to make the runtime exception disappear. The main goal is to have non-zero values in column ``\# Valid''.}
  \label{tab:generation}
  \begin{tabularx}{0.9\textwidth}{|X|r|r|r|r|}
    \hline
    & \multicolumn{4}{c|}{Patch Models} \\\cline{2-5}
    Bug &
    \multicolumn{2}{c|}{NPEFix} & \multicolumn{2}{c|}{Exception-Stopper} \\\cline{2-5}
    & \# Valid & \# Invalid & \# Valid & \# Invalid \\
    \hline
    BroadleafCommerce 1282 & 5  & 8 &  0   &  0 \\
    Collection 360 & 16  & 35 &  64   &  44 \\
    DataflowJavaSDK c06125d & 2  & 1 &  0   &  0 \\
    Felix 4960 & 4  & 6 &  0   &  1 \\
    Javapoet 70b38e5 & 0  & 133 &  12   &  87 \\
    Jetty 335500 & 2  & 2 &  2   &  2 \\
    Jongo f46f658 & 0  & 1 &  2   &  8 \\
    Lang 20 & 78  & 634 &  0   &  15 \\
    Lang 304 & 65  & 12 &  32   &  276 \\
    Lang 33 & 1  & 27 &  0   &  3 \\
    Lang 39 & 4  & 7 &  0   &  8 \\
    Lang 587 & 28  & 0 &  3   &  0 \\
    Lang 703 & 7  & 12 &  0   &  15 \\
    Math 1115 & 5  & 6 &  4   &  1 \\
    Math 1117 & \numprint{22132}  & \numprint{29711} &  0   &  0 \\
    Math 290 & 4  & 10 &  4   &  3 \\
    Math 305 & 3  & 1 &  9   &  1 \\
    Math 369 & 23  & 3 &  22   &  2 \\
    Math 4 & 415  & 95 &  2   &  17 \\
    Math 70 & 1  & 25 &  0   &  24 \\
    Math 79 & 0  & 4 &  0   &  10 \\
    Math 988A & 168  & 37 &  8   &  11 \\
    Math 988B & 17  & 15 &  1   &  12 \\
    Mayocat 231 & 102  & 182 & 18 & 19 \\
    PDFBox 2812 & 4  & 21 &  2   &  9 \\
    PDFBox 2965 & 3  & 1 &  1   &  0 \\
    PDFBox 2995 & 1  & 4 &  1   &  1 \\
    Sling 4982 & 11  & 9 &  6   &  4 \\
    Tomcat 43758 & 1  & 9 &  1  &  0 \\
    Tomcat 54703 & 10  & 0 &  2  &  1 \\
    Tomcat 55454 & 1  & 0 &  1  &  0 \\
    Tomcat 56010 & 0  & 0 & 0 & 7 \\
    Tomcat 58232 & 3  & 0 &  0   &  0 \\
    Webmagic ff2f588 & 2  & 49 &  0   &  10 \\
    \hline
    {34 bugs from 14 software applications} 
    & \numprint{23118} 
    & \numprint{31060} 
    & 198 
    & 592 \\
    \hline
  \end{tabularx}
\end{table*}

\subsubsection{Benchmark}\label{sec:generation-benchmark}

In order to evaluate whether patch generation can be made directly on production failures, we need to identify real reproducible failures.
To collect as many production bugs as possible, we build a benchmark based on the failures used in five different papers from the literature:
\cite{JustJE2014}, \cite{long2017automatic}, \cite{gu2016automatic}, \cite{npefix} and  \cite{DurieuxHM16}.

Our inclusion criteria are as follows.
First, we select the exception bugs. An exception bug is an unhandled exception in production which makes a request crash.
Second, we only keep the bugs for which we are able to replay user traffic or setup that triggers the bug.
Third, we discard the failures that happen during initialization or shutdown of an application.

By applying the inclusion criteria, we eventually come up with \generationNbBug real production failures from 14 different applications. 
The benchmark contains 33 null pointer exceptions and one invalid argument exception.
For sake of open-science, this benchmark is publicly available on GitHub \cite{repo}.

In \autoref{tab:bench-generation}, the first five columns present the dataset.
The first column contains a simple bug identity, 
the second column contains the origin of the bug,
the third column contains the type of production oracle considered, 
the fourth column contains the type of the failure (NPE for Null Pointer Exception, IAE for IllegalArgumentException),
and the fifth column contains the number of lines of Java code of the buggy application under consideration, which is computed with the CLOC tool. 

\subsubsection{Experimental Protocol for RQ1}\label{sec:generation-protocol}

To evaluate the patch generation of \itzal, we set up the following experimental protocol.
The main idea of this experimental protocol is to execute a HTTP request that triggers a failure.
Based on this failure the \patchService searches for patches.
The main novelty of this setup compared to test-suite based program repair is the following: while previous experiments assume a manually written failing test case, \itzal only assumes a failing request. This is greatly widens the applicability of the approach.

For the bug of our benchmark that uses HTTP status as failure oracle, we simulate a server that runs the buggy version. 
This server waits for requests, as a production server would do.
Then, we send a request that triggers the failure.
We check whether the failure is well detected by the \requestOracle, the HTTP status in this case.
Then, we put the failure-triggering request in an infinite loop to simulate arriving user requests that trigger the  failure, making the same failure happening again and again, as in production.
Each time the failure happens, it triggers a patch search by the \patchService.
Hence, the \patchService enumerates all candidate patches and identify those that make the failure disappear, i.e., that pass the \requestOracle. 

For the other bugs due to unhandled exceptions, we encapsulate a small execution scenario that triggers the unhandled exception into a HTTP request that could be run again and again. This small execution scenario is also put in an infinite loop, as what happens in production with user-generated requests.

Finally, we use the two patch models described in \autoref{sec:implem:path-search} to generate patches for the bugs with. We count the number of invalid and valid patches for each failure and for each patch model.

\subsubsection{Results}\label{sec:generation-results}
We now present the results for this research question. To answer this question, we consider the columns \emph{\# NPEFix Valid/Invalid} and \emph{\# Exception-Stopper Valid/Invalid} of \autoref{tab:generation}, which show the number of valid and invalid patches generated by our two patch models respectively. Valid means the initial failure does not happen anymore, and no other exceptions happen. Invalid means that the initial failure still happens or other failures happen.
The main goal is to have non-zero values in column ``\# Valid'', this shows the feasibility of our vision.

For example, the first row of \autoref{tab:generation} presents the result for bug BroadleafCommerce 1282.
This bug is caused by a null dereference happening upon a user request.
We assert the presence of the bug in the application by using an HTTP-based production oracle: HTTP status.
The first repair model, NPEFix, generates 13 candidate patches, including 5 valid patches and 8 invalid patches.
The second repair model, Exception-Stop, does not generate any patch for this specific bug.

Overall, we can see from \autoref{tab:generation} that it is possible to generate patches for real-life production failures. 
For all the 34 failures of our benchmark, at least one patch can be generated by the two patch models used by \itzal.
The \requestOracle is capable of discarding many invalid patches that are in the search space of the considered synthesis techniques.

The number of generated patches varies significantly between projects and failures, the number of candidate patches ranges from 2 (for Tomcat 55454) to 51843 (for Math 1117) and the number of valid patches varies between 0 (for several failures) and 22132 (for Math 1117).
This difference in the number of generated patches from the two patch models emerges as NPEFix is able to generate more patches than Exception Stopper in general.
The underlying reason is that the search space of the Exception Stopper patch model is smaller than that of NPEFix.
The search space of Exception Stopper is defined by the number of method calls in the stack and the number of variable/value pairs that are available for returning from the current method. 
Instead, the search space of NPEFix is defined by 9 repair strategies that contain several variants (different values for the placeholder) depending on the context of execution.

Interestingly, we can see from the table that there are a lot of valid patches for both considered patch models. This is a challenge because one would obviously not report to the developer so many patches. However, this issue is handled later in \itzal because 1) the \regressionService further removes patches and 2) the patches are displayed to the developers by the order of potential value, as discussed in \autoref{sec:dashboard}.

Meanwhile, we can also see from the table that NPEFix also has proportionally more valid patches than Exception Stopper.
This can be explained by two facts. On the one hand, Exception Stopper is a generic repair technique, which works at the coarse-grain level. But NPEfix works at the point-of-failure (statement level), and thus it generates patches that are more likely to be invalid.
On the other hand, our benchmark contains mostly null dereferences. NPEFix is thus favored as its strategies are specifically designed to handle such failures.
Note NPEFix is unable to handle failures that are not null dereference by design (this is what happens for the failure in Tomcat 56010).

To sum up, for all the \generationNbBug failures of our benchmark, we show that it is possible to generate patches using one or both patch models implemented in \itzal.
This is a large proof-of-concept that it is possible to generate patches directly from production failures.

\answer{1}{This novel experiment on \generationNbBug production failures shows that one can replay a failing request to explore the search space of a patch model.
Thus, it is possible to generate patches directly in production based on user traffic. Our experimentation with two different patch models shows that \itzal is oblivious to the actual patch generation technique.}

\subsection{RQ2. User Traffic Effectiveness for Regression.}
\label{sec:rq2}
\begin{table}
  \centering
  \caption{The Effectiveness of four Execution Comparison Oracles to Detect Regressions based on User Traffic. A green plain circle means that the oracle is effective at detecting the regression.}
  \label{tab:regression}
  \resizebox*{!}{\dimexpr\textheight-2.9\baselineskip\relax}{
  \begin{tabular}{|l|l|r|r|r|r|r|}
    \hline
    & & \multicolumn{4}{c|}{Oracles} & 
    \multirow{ 2}{*}{\rotatebox{0}{Is Valid Patch}} \\ \cline{3-6}
    \rotatebox{0}{Projects} & 
    Patch Location  & 
    \rotatebox{0}{HTTP status} & 
    \rotatebox{0}{HTTP content} & 
    \rotatebox{0}{\# Method} & 
    \rotatebox{0}{\# Block} & \\
    \hline 
    \multirow{ 17}{*}{\rotatebox{0}{Broadleaf}}
    & CategoryImpl:835 1 
    & 0\%  \failing 
    & 17\%  \passing 
    & 36\% \passing 
    & 40\% \passing  
    & No \\
    & CategoryImpl:835 2
    & 0\%  \failing 
    & 17\%  \passing 
    & 36\% \passing 
    & 40\% \passing  
    & No \\
    & CategoryImpl:835 3
    & 0\%  \failing 
    & 20\%  \passing 
    & 36\% \passing 
    & 41\% \passing  
    & No \\
    & CategoryImpl:835 4
    & 0\%  \failing 
    & 17\%  \passing 
    & 36\% \passing 
    & 40\% \passing  
    & No \\
    & CategoryImpl:835 5
    & 0\%  \failing 
    & 20\%  \passing 
    & 36\% \passing 
    & 41\% \passing  
    & No \\
    & CategoryImpl:835 6
    & 0\%  \failing 
    & 20\%  \passing 
    & 36\% \passing 
    & 41\% \passing  
    & No \\
    & CategoryImpl:835 7 
    & 45\%  \passing 
    & 48\%  \passing 
    & 42\% \passing 
    & 47\% \passing  
    & No \\
    & OrderItemImpl:418 1 \validPatch
    & 0\%  \failing 
    & 1\%  \passing 
    & 0\% \failing 
    & 0\% \failing  
    & Yes \\
    & OrderItemImpl:418 2 
    & 0\%  \failing 
    & 1\%  \passing 
    & 0\% \failing 
    & 0\% \failing  
    & No \\
    & OrderItemImpl:418 3 \validPatch
    & 0\%  \failing 
    & 1\%  \passing 
    & 0\% \failing 
    & 0\% \failing  
    & Yes \\
    & RelatedProductsServiceImpl:208 1 \validPatch
    & 0\%  \failing 
    & 1\%  \passing 
    & 0\% \failing 
    & 0\% \failing  
    & Yes \\
    & RelatedProductsServiceImpl:208 2
    & 0\%  \failing 
    & 7\%  \passing 
    & 34\% \passing 
    & 38\% \passing  
    & No \\
    & RelatedProductsServiceImpl:208 3
    & 0\%  \failing 
    & 15\%  \passing 
    & 3\% \passing 
    & 4\% \passing  
    & No \\
    & SolrHelperServiceImpl:531 1
    & 0\%  \failing 
    & 13\%  \passing 
    & 0\% \failing 
    & 0\% \failing  
    & No \\
    & SolrHelperServiceImpl:531 2
    & 35\%  \passing 
    & 37\%  \passing 
    & 11\% \passing 
    & 9\% \passing  
    & No \\
    & SolrHelperServiceImpl:531 3
    & 35\%  \passing 
    & 37\%  \passing 
    & 11\% \passing 
    & 9\% \passing  
    & No \\
    & SolrHelperServiceImpl:531 4
    & 35\%  \passing 
    & 37\%  \passing 
    & 11\% \passing 
    & 9\% \passing  
    & No \\
    & SolrHelperServiceImpl:531 5
    & 35\%  \passing 
    & 37\%  \passing 
    & 11\% \passing 
    & 9\% \passing  
    & No \\
    \hline
    \multirow{39}{*}{\rotatebox{0}{Mayocat}}
    & AbstractScopeCookieContainerFilter:202 1 \validPatch
    & 0\%  \failing 
    & 0\%  \failing 
    & 0\% \failing 
    & 0\% \failing  
    & Yes \\
    & AbstractScopeCookieContainerFilter:202 2 \validPatch
    & 0\%  \failing 
    & 0\%  \failing 
    & 0\% \failing 
    & 0\% \failing  
    & Yes \\
    & AbstractScopeCookieContainerFilter:202 3
    & 0\%  \failing 
    & 21\%  \passing 
    & 6\% \passing 
    & 6\% \passing  
    & No \\
    & AbstractScopeCookieContainerFilter:202 4
    & 0\%  \failing 
    & 21\%  \passing 
    & 6\% \passing 
    & 6\% \passing  
    & No \\
    & AbstractScopeCookieContainerFilter:202 5
    & 0\%  \failing 
    & 21\%  \passing 
    & 6\% \passing 
    & 6\% \passing  
    & No \\
    & AbstractScopeCookieContainerFilter:256 1 \validPatch
    & 0\%  \failing 
    & 0\%  \failing 
    & 0\% \failing 
    & 0\% \failing  
    & Yes \\
    & AbstractScopeCookieContainerFilter:256 2
    & 0\%  \failing 
    & 0\%  \failing 
    & 0\% \failing 
    & 0\% \failing  
    & No \\
    & AbstractScopeCookieContainerFilter:256 3
    & 0\%  \failing 
    & 0\%  \failing 
    & 0\% \failing 
    & 0\% \failing  
    & No \\
    & AbstractScopeCookieContainerFilter:256 4
    & 0\%  \failing 
    & 0\%  \failing 
    & 0\% \failing 
    & 0\% \failing  
    & No \\
    & DateAsTimestampArgumentFactory:30 1 \validPatch
    & 0\%  \failing 
    & 0\%  \failing 
    & 0\% \failing 
    & 0\% \failing  
    & Yes \\
    & DateAsTimestampArgumentFactory:30 2
    & 0\%  \failing 
    & 0\%  \failing 
    & 1\% \passing 
    & 1\% \passing  
    & No \\
    & DateAsTimestampArgumentFactory:30 3 \validPatch
    & 0\%  \failing 
    & 0\%  \failing 
    & 0\% \failing 
    & 0\% \failing  
    & Yes \\

    & DateAsTimestampArgumentFactory:30 4
    & 0\%  \failing 
    & 0\%  \failing 
    & 0\% \failing 
    & 0\% \failing  
    & No \\

    & DateAsTimestampArgumentFactory:30 5
    & 0\%  \failing 
    & 0\%  \failing 
    & 0\% \failing 
    & 0\% \failing  
    & No \\

    & DefaultCartLoader:88 1
    & 82\%  \passing 
    & 82\%  \passing 
    & 18\% \passing 
    & 16\% \passing  
    & No \\

    & DefaultCartLoader:88 2 \validPatch
    & 0\%  \failing 
    & 0\%  \failing 
    & 0\% \failing 
    & 0\% \failing  
    & Yes \\

    & DefaultCartManager:198 1 \validPatch
    & 0\%  \failing 
    & 0\%  \failing 
    & 0\% \failing 
    & 0\% \failing  
    & Yes \\

    & DefaultCartManager:198 2
    & 0\%  \failing 
    & 0\%  \failing 
    & 0\% \failing 
    & 0\% \failing  
    & No \\

    & FlatStrategyPriceCalculator:38 1 \validPatch
    & 0\%  \failing 
    & 0\%  \failing 
    & 0\% \failing 
    & 0\% \failing  
    & Yes \\

    & FlatStrategyPriceCalculator:38 2
    & 0\%  \failing 
    & 5\%  \passing 
    & 0\% \failing 
    & 0\% \failing  
    & No \\

    & FlatStrategyPriceCalculator:38 3 \validPatch
    & 0\%  \failing 
    & 0\%  \failing 
    & 0\% \failing 
    & 0\% \failing  
    & Yes \\

    & FlatStrategyPriceCalculator:38 4
    & 0\%  \failing 
    & 5\%  \passing 
    & 0\% \failing 
    & 0\% \failing  
    & No \\

    & FlatStrategyPriceCalculator:38 5
    & 20\%  \passing 
    & 20\%  \passing 
    & 1\% \passing 
    & 1\% \passing  
    & No \\

    & FlatStrategyPriceCalculator:38 6 \validPatch
    & 0\%  \failing 
    & 0\%  \failing 
    & 0\% \failing 
    & 0\% \failing  
    & Yes \\

    & FlatStrategyPriceCalculator:38 7
    & 20\%  \passing 
    & 20\%  \passing 
    & 1\% \passing 
    & 1\% \passing  
    & No \\

    & MapAsJsonArgumentFactory:30 1 \validPatch
    & 0\%  \failing 
    & 0\%  \failing 
    & 0\% \failing 
    & 0\% \failing  
    & Yes \\

    & MapAsJsonArgumentFactory:30 2
    & 0\%  \failing 
    & 0\%  \failing 
    & 2\% \passing 
    & 2\% \passing  
    & No \\

    & MapAsJsonArgumentFactory:30 3
    & 0\%  \failing 
    & 0\%  \failing 
    & 0\% \failing 
    & 0\% \failing  
    & No \\

    & MapAsJsonArgumentFactory:30 4 \validPatch
    & 0\%  \failing 
    & 0\%  \failing 
    & 0\% \failing 
    & 0\% \failing  
    & Yes \\

    & MapAsJsonArgumentFactory:30 5 \validPatch
    & 0\%  \failing 
    & 0\%  \failing 
    & 0\% \failing 
    & 0\% \failing  
    & Yes \\

    & PostgresUUIDArrayArgumentFactory:30 1 \validPatch
    & 0\%  \failing 
    & 0\%  \failing 
    & 1\% \passing 
    & 2\% \passing  
    & Yes \\

    & PostgresUUIDArrayArgumentFactory:30 2
    & 0\%  \failing 
    & 0\%  \failing 
    & 0\% \failing 
    & 1\% \passing  
    & No \\

    & PostgresUUIDArrayArgumentFactory:30 3 \validPatch
    & 0\%  \failing 
    & 0\%  \failing 
    & 0\% \failing 
    & 1\% \passing  
    & Yes \\

    & PostgresUUIDArrayArgumentFactory:30 4 \validPatch
    & 0\%  \failing 
    & 0\%  \failing 
    & 0\% \failing 
    & 1\% \passing  
    & Yes \\

    & ProductMapper:44 1 \validPatch
    & 0\%  \failing 
    & 0\%  \failing 
    & 0\% \failing 
    & 0\% \failing  
    & Yes \\

    & ProductMapper:44 2
    & 0\%  \failing 
    & 0\%  \failing 
    & 0\% \failing 
    & 0\% \failing  
    & No \\

    & ProductMapper:44 3
    & 0\%  \failing 
    & 0\%  \failing 
    & 0\% \failing 
    & 0\% \failing  
    & No \\

    & ProductMapper:44 4
    & 0\%  \failing 
    & 0\%  \failing 
    & 0\% \failing 
    & 0\% \failing  
    & No \\

    & ProductMapper:44 5 
    & 0\%  \failing 
    & 0\%  \failing 
    & 0\% \failing 
    & 0\% \failing  
    & No \\
    \hline
    \multirow{ 22}{*}{\rotatebox{0}{Shopizer}}
    & CategoryFacadeImpl:55 1 \validPatch
    & 0\%  \failing 
    & 0\%  \failing 
    & 0\% \failing 
    & 0\% \failing  
    & Yes \\
    & CategoryFacadeImpl:55 2
    & 0\%  \failing 
    & 67\%  \passing 
    & 6\% \passing 
    & 5\% \passing  
    & No \\
    & CategoryFacadeImpl:55 3
    & 0\%  \failing 
    & 67\%  \passing 
    & 18\% \passing 
    & 18\% \passing  
    & No \\
    & CategoryFacadeImpl:55 4
    & 0\%  \failing 
    & 67\%  \passing 
    & 15\% \passing 
    & 14\% \passing  
    & No \\
    & CategoryFacadeImpl:55 5
    & 0\%  \failing 
    & 67\%  \passing 
    & 18\% \passing 
    & 18\% \passing  
    & No \\
    & ReadableCategoryPopulator:51 1
    & 0\%  \failing 
    & 0\%  \failing 
    & 1\% \passing 
    & 1\% \passing  
    & No \\
    & ReadableCategoryPopulator:51 2
    & 0\%  \failing 
    & 67\%  \passing 
    & 4\% \passing 
    & 3\% \passing  
    & No \\
    & ReadableCategoryPopulator:51 3
    & 12\%  \passing 
    & 67\%  \passing 
    & 10\% \passing 
    & 7\% \passing  
    & No \\
    & ReadableCategoryPopulator:51 4
    & 0\%  \failing 
    & 67\%  \passing 
    & 6\% \passing 
    & 4\% \passing  
    & No \\
    & ReadableCategoryPopulator:51 5
    & 0\%  \failing 
    & 67\%  \passing 
    & 6\% \passing 
    & 5\% \passing  
    & No \\
    & ReadableProductPopulator:94 1 \validPatch
    & 0\%  \failing 
    & 0\%  \failing 
    & 0\% \failing 
    & 0\% \failing  
    & Yes \\
    & ReadableProductPopulator:94 2 
    & 0\%  \failing 
    & 0\%  \failing 
    & 0\% \failing 
    & 0\% \failing  
    & No \\
    & ReadableProductPopulator:94 3
    & 26\%  \passing 
    & 37\%  \passing 
    & 9\% \passing 
    & 6\% \passing  
    & No \\
    & ReadableProductPopulator:94 4
    & 26\%  \passing 
    & 37\%  \passing 
    & 11\% \passing 
    & 7\% \passing  
    & No \\
    & ReadableProductPopulator:94 5
    & 26\%  \passing 
    & 37\%  \passing 
    & 11\% \passing 
    & 7\% \passing  
    & No \\
    & ShoppingCategoryController:253 1 \validPatch
    & 0\%  \failing 
    & 0\%  \failing 
    & 0\% \failing 
    & 0\% \failing  
    & Yes \\

    & ShoppingCategoryController:253 2 
    & 0\%  \failing 
    & 0\%  \failing 
    & 0\% \failing 
    & 0\% \failing  
    & No \\

    & ShoppingCategoryController:253 3
    & 0\%  \failing 
    & 0\%  \failing 
    & 0\% \failing 
    & 0\% \failing  
    & No \\

    & ShoppingCategoryController:253 4
    & 0\%  \failing 
    & 12\%  \passing 
    & 2\% \passing 
    & 1\% \passing  
    & No \\

    & ShoppingCategoryController:253 5
    & 12\%  \passing 
    & 12\%  \passing 
    & 2\% \passing 
    & 1\% \passing  
    & No \\

    & ShoppingCategoryController:253 6
    & 12\%  \passing 
    & 12\%  \passing 
    & 2\% \passing 
    & 1\% \passing  
    & No \\

    & ShoppingCategoryController:253 7
    & 12\%  \passing 
    & 12\%  \passing 
    & 2\% \passing 
    & 1\% \passing  
    & No \\

    & ShoppingCategoryController:253 8
    & 12\%  \passing 
    & 12\%  \passing 
    & 2\% \passing 
    & 1\% \passing  
    & No \\
    \hline
    & 80 patches from 17 locations & 16 & 42 & 39 & 42 & 23 valid patches\\
    \hline
  \end{tabular}
  }
\end{table}

We have shown that it is possible to generate patches directly from user traffic.
We are now interested in seeing if it is possible to use user traffic to discard regressive patches.

\subsubsection{Benchmark}\label{sec:regression-benchmark}
The benchmark of RQ1 has a single request, i.e., the failure-triggering one.
For this second research question, we need several requests for the same application, i.e., a workload.
We search for HTTP applications on the GitHub software repository with a focus on e-commerce applications as e-commerce applications are easy to understand and consequently, we can create a meaningful workload. 

We identify three e-commerce applications that meet our criteria: Mayocat\footnote{\url{http://www.mayocat.org/}}, BroadLeaf Commerce\footnote{\url{http://www.broadleafcommerce.com/}}, and Shopizer\footnote{\url{http://www.shopizer.com/}}.
Mayocat is composed of \numprint{31231} lines of Java code, done over \numprint{1670} commits, and in development since 2012.
BroadleafCommerce is bigger, it is composed of \numprint{154309} lines of code, done over \numprint{9779} commits, and in development since 2008.
Shopizer is composed of \numprint{61555} lines of Java code, done over \numprint{154} commits, and in development since 2015.
Similarly, for the sake of open science, this benchmark is made publicly available on GitHub \cite{repo}. 

\emph{User traffic:} For each of these three e-commerce applications, we create a user traffic by identifying a set of requests that execute the major user features, such as adding an item to the cart.
Then we automatically create 25 different user sessions that contain between 3 and 7 requests, selected randomly from our set of requests. For sake of reproducibility, we always use the same random seed.
Consequently, we generate a user traffic of 124 requests for each e-commerce application.
By keeping the number of requests below 200, the experimental time remains manageable.

\emph{Failures:} 
Since we aim at studying the \regressionService which detects regressions introduced by patches, we need such patches.
To achieve this, we first seed faults into the programs and then consider a sample of patches generated by the patch model under consideration for the seeded faults.
We seed null dereference faults by removing the ``then'' block of a randomly sampled not-null check that has been executed.
For example, if an executed not-null check is ``if (x==null) then A else B'', we rewrite it as ``B''.
In other words, we remove the error-handling code which deals with null values. We further check whether the seeded faults really trigger failures. Eventually, we have 16 seeded faults that trigger failures under our emulated user traffic. 

\emph{Patches:} 
For each seeded fault, we select a random sample of patches that are in the search space of NPEFix, one of the patch models implemented in the prototype implementation of \itzal. 
We obtain a benchmark of 80 candidate patches to be considered for regression. 
The first two columns of \autoref{tab:regression} present this benchmark.

\subsubsection{Experimental Protocol for RQ2}\label{sec:regression-protocol}

To evaluate the \regressionService, we first execute the user traffic as described in \autoref{sec:regression-benchmark} on each considered patch. 

Then, for each request of the user traffic, we compare the execution of the patched application against that of the original application. 
In this context, it means defining a point of observation, collecting some values at this point, and comparing the values collected on the original application with that collected on the patched application. This enables us to observe differences, called ``divergence'' in the rest of this paper, following the terminology introduced in \cite{palikareva2016shadow}.
If there exists a divergence for a normal successful request, the patch is considered as a regression. 

In this experiment, we consider four different execution comparison oracles to capture divergences. The first is the HTTP status of the response, the second is the HTTP content of the response, the third is the set of covered methods, and the final one is the set of covered blocks. For HTTP status and HTTP content, we collect the percentage of requests for which we observe differences in HTTP status and content respectively. For method and block coverage, we first collect the coverage divergence for each request and then compute the average value over all requests.
Finally, we compare the oracle results against a manual analysis of the generated patches.

\subsubsection{Results}\label{sec:regression-results}

\autoref{tab:regression} contains the data obtained with the experimental protocol described in \autoref{sec:regression-protocol}, investigating whether the execution comparison oracles considered are able to detect divergences.
The first column of \autoref{tab:regression} contains the name of the project under consideration.
The second column contains the patch id which is composed of the class name, the line number, and a sequential id. 
The four next columns of \autoref{tab:regression} provide the measured divergence between the original program and the seeded programs using the  4 execution comparison oracles presented in \autoref{sec:regression-protocol}.
A green plain circle \passing means that the considered comparison oracle is able to detect a regression on at least one request, which is desirable in the context of patch generation in production.
A crossed circle \failing means that the execution comparison oracle fails to detect a divergence.
An ideal oracle would detect all divergences, but this would require to compare the whole execution state which is impossible in practice.
The last column, Is valid Patch, indicates if the generate patch is semantically correct according to our manual analysis.

For example, the first row of \autoref{tab:regression} describes a patch for Broadleaf at line 835 of file CategoryImpl. For this patch, the HTTP status does not detect a single divergence, the HTTP content detects a divergence for 17\% of the requests (i.e., 17\% of the requests have different contents compared to the original program), and the average divergences of method and block coverage across all requests are 36\% and 40\% respectively.
When a line contains 0\% \failing for all four oracles, it means that either the patch is correct (hence has no regressions) or that the synthetic workload we use is not rich enough to highlight the regressions.

The HTTP status execution comparison oracle is easy to obtain but is at a relatively coarse level. Indeed, we can see from \autoref{tab:regression} that the HTTP status oracle discards only 16 patches of the whole 80 patches. Since we do not have a reliable correctness oracle, it is not meaningful to compute information retrieval metrics such as precision and recall. 
The HTTP status execution comparison oracle has two main advantages. On the one hand, it has virtually no overhead. On the other hand, it is directly generalizable to any HTTP based applications.

The HTTP content execution comparison oracle detects regressions for 42 of the whole 80 patches, which is better than the HTTP status execution comparison oracle.
However, it has a drawback:
it requires to define some transformations on the output in order to remove transient information. In \itzal, the response is cleaned by removing certain transient information, e.g., date, a session key or dynamic CSS classes.
It is not always possible to identify all transient information. 
For instance, for the patches at line 418 of file OrderItemImpl for project Broadleaf, the HTTP content execution comparison oracle is considered as regressive because random inconsistencies happen in the HTML response (one letter disappears at different locations). In the context of patch generation in production, this means that some patches would be discarded because of transient information but not because of an actual regression.

The method coverage and block coverage execution comparison oracles both detect more regressions and have almost the same behavior. Compared to method coverage execution comparison oracle, the block coverage execution comparison oracle detects regressions for 3 more patches located at file PostgresUUIDArrayArgumentFactory for project Mayocat. While these two execution comparison oracles are effective at detecting behavior changes, we have observed an issue: parallel execution can introduce some randomness and consequently some variance in the observed dynamic coverage for a given request. This can possibly be a reason for having correct patches that are discarded by at least one oracle.

We consider the HTTP content as the best execution comparison oracle for regression detection based on user traffic. The reasons are:
1) it is quite effective at detecting behavioral changes;
2) its sensitivity can be overcome with careful design (on the contrary, it is virtually impossible to overcome the non-determinism of observed coverage due to concurrency).

Now let us discuss the oracle results against the actual correctness as found by manual analysis.
Our manual analysis of the generated patches reveals that 16 patches are incorrect. However, these 16 incorrect patches have not been detected by any of the four oracles. 
In other words, the rows for these 16 patches are indicated as invalid but with \failing for all the 4 oracles in \autoref{tab:regression}, e.g., AbstractScopeCookieContainerFilter:256.
The reason of this phenomenon is that our simulated HTTP workload is not able to produce inputs that trigger the invalid behaviors of the incorrect patches.
There are also 6 patches for which the opposite phenomenon occurs: they are correct but they are discarded due to randomness and multithreading as discussed above.

\answer{2}{
It is possible to to employ user traffic to validate patches. 
To substantiate this claim, our novel experimental methodology compares different execution comparison oracles that are all available in production.
In the context of HTTP based applications, we observe that the HTTP status oracle, HTTP content oracle, method coverage  oracle and block coverage  oracle successfully discard 16, 52, 39 and 42 out of 80 patches respectively. This result shows that 1) the \itzal novel scheme is generic enough to accommodate different execution comparison oracles; 2) using an HTTP content based execution comparison oracle represents a good trade-off to perform live regression testing in message-driven applications.}

\subsection{RQ3. \itzal Performance}
\label{sec:rq3}

We now consider the performance of \itzal.
We will focus first on the impact of the \shadowService on the performance of the application, and then evaluate how much time \itzal requires to generate patches.

\subsubsection{\shadowService Overhead}

As we have previously discussed in \autoref{sec:shadow}, the performance of the application is only impacted by the \shadowService.
Since the other two services (\patchService and \regressionService) are executed asynchronously (no overhead on the response time), the \shadowService is the only one that may have user-visible impact.

In order to evaluate the performance impact of \shadowService on the application and further on the clients, we create a workload of 100 000 requests.
We compare the performance of those requests with the \shadowService and without the \shadowService.
First, these 100 000 requests are launched sequentially on the  Mayocat application without the \shadowService.  
Second, we execute the same 100 000 requests but this time with the \itzal \shadowService.
We collect the average response time for the two infrastructures.

We observe that it takes on average 104ms to make a request directly to Mayocat. 
With \itzal \shadowService, it takes on average 114ms to make a request through the \shadowService.
This represents a slowdown of 10ms per request or an overhead of 10.44\% on average. 
The reason for the slowdown is that the \shadowService costs some time to copy the request to \patchService and \regressionService, redirect the original request to the Mayocat application, and finally copy the response of Mayocat to the \regressionService.

\subsubsection{\patchService Performance}

The role of the \patchService is to generate the patches for the failing request. We now study how much time the \patchService needs to generate these patches.

We apply \itzal on the bug Mayocat-231 and we execute the request that produces the failure.
We measure how much time the \patchService takes to exhaustively generate the patches with the NPEFix repair model for this bug.

The result shows that the \patchService takes 4min and 33sec to generate the 286 \canditatePatches, which means that the \patchService takes on average 953ms to generate one patch.
Recall that this has no impact on the user, because  the \patchService is called in an asynchronous manner by the \shadowService (see \autoref{sec:shadow}). In other words, the end-user does not have to wait for 5 minutes in front of the browser.

\answer{3}{
By design, the only component of \itzal that has an overhead in production is the \shadowService (all other components being called asynchronously, with no blocking callbacks).
Our experimental evaluation the \shadowService's overhead, 
shows that it adds on average a 10ms latency per client request, which is negligible from a user experience perspective.
When a failure is detected, according to our benchmark, \itzal is able to generate one per patch per second, and the time to explore the search space is linear in the number of patches.}

\subsection{RQ4. End-to-end Effectiveness of \itzal}
\label{sec:rq4}

While research questions RQ1, RQ2, and RQ3 concentrate on evaluating the \patchService, \regressionService, and \shadowService, the fourth research question aims to evaluate \itzal from an end-to-end perspective by considering two real bugs that are reproducible in a live server environment with emulated traffic.
We do the end-to-end evaluation on two real bugs from open-source ecommerce applications--Mayocat-231 and BroadleafCommerce-1282. Mayocat-231 is our working case and has already been used in all the above three research questions, and BroadleafCommerce-1282 has been already used in RQ1.

\subsubsection{Experimental Protocol for RQ4}\label{sec:endtoend-protocol}

To evaluate the end-to-end effectiveness of \itzal, we apply \itzal on the two applications.
The infrastructure consists of four different docker images and an instance of \shadowService that duplicates the requests and the responses to the different services.
The first docker image contains the \productionApplication with the correct state to reproduce the bug.
The second and the third docker images are for the \patchService (one per patch model NPEFix or Exception-stopper).
The last docker image contains the \regressionService, the infrastructure to identify regressions due to the generated patches.

Once the infrastructure is up and running, we launch the failing HTTP request that triggers the patch generation by the \patchService.
Then, each generated patch will be evaluated by the \regressionService against the emulated traffic. Note that the emulated traffic here is the same as the traffic used in the experiment for RQ2,

\subsubsection{End-to-end Evaluation on Mayocat-231}

\begin{figure}[t]
\centering
\begin{lstlisting}[language=diff]
@@ FlatStrategyPriceCalculator.java
@@ -35,7 +35,8 @@
        return BigDecimal.ZERO;
    }
-   price = price.add(carrier.getPerItem().multiply(BigDecimal.valueOf(numberOfItems)));
+   BigDecimal perItem = carrier.getPerItem() != null ? carrier.getPerItem() : BigDecimal.ZERO;
+   price = price.add(perItem.multiply(BigDecimal.valueOf(numberOfItems)));
    return price;
 }
\end{lstlisting}
\caption{The human patch for bug Mayocat 231.}
\label{lst:Mayocat-231:patch}
\end{figure}

\begin{figure}[t]
\centering
\begin{lstlisting}[language=diff]
@@ FlatStrategyPriceCalculator.java
@@ -37,2 +37,5 @@
+   if (carrier.getPerItem() == null) {
+      return null;
+   }
    price = price.add(carrier.getPerItem().multiply(BigDecimal.valueOf(numberOfItems)));
\end{lstlisting}
\caption{An invalid \itzal patch for bug Mayocat 231.}
\label{lst:Mayocat-231:invalid_patch}
\end{figure}

\begin{figure}[t]
\centering
\begin{lstlisting}[language=diff]
@@ FlatStrategyPriceCalculator.java
@@ -37,3 +37,7 @@
-   price = price.add(carrier.getPerItem().multiply(BigDecimal.valueOf(numberOfItems)));
+   if (carrier.getPerItem() == null) {
+      price = price.add( BigDecimal.ZERO.multiply(BigDecimal.valueOf(numberOfItems)));
+   } else {
+      price = price.add(carrier.getPerItem().multiply(BigDecimal.valueOf(numberOfItems)));
+   }
    return price;
\end{lstlisting}
\caption{A patch found by \itzal for bug Mayocat 231.}
\label{lst:Mayocat-231:itzal_patch}
\end{figure}

\paragraph{Description of the bug}

This bug is an unhandled null pointer exception of the e-commerce application Mayocat (\url{https://github.com/jvelo/mayocat-shop/issues/231}).
The bug is triggered during the computation of the shipping cost of the current cart.
This bug is present only for one specific shipping strategy.
When the bug happens, the user is left with a white page. Worse still, the user session becomes completely unusable, which means that the website is completely broken for this particular user. 
The client is thus unable to further navigate through the product list, buy a product or even click on the ``contact the administrator'' link to report the issue.

\paragraph{Human patch}

\autoref{lst:Mayocat-231:patch} shows the snippet of code written by the human developer to fix the bug.
The patch consists of using ``BigDecimal.ZERO'' when the shipping price per product (``carrier.getPerItem()'') is null. It is a classical patch for null dereferences: adding a not-null check, here in the form of a ternary expression.

\paragraph{\patchService}

For the failing request, the \patchService of \itzal generates 284 \canditatePatches with patch model NPEFix  and 37 \canditatePatches with patch model Exception-Stopper  (the complete list of the \itzal patches is available at \cite{repo}).
Out of the 321 (284 + 37) \canditatePatches, 201 (182 + 19) fail to make the exception disappear or produce another exception so that they are considered as invalid (as we did in RQ1).
For example, let us consider the candidate patch shown in \autoref{lst:Mayocat-231:invalid_patch}.
This patch is invalid according to the \requestOracle, because it produces a HTTP status 500, i.e., it indicates an internal server error. 
The reason is that when this patch is applied, a null value is returned, which itself produces a new null pointer exception in the caller method. 
This new null pointer exception makes the request fail and the server eventually returns a HTTP 500 code.
The \requestOracle well detects the HTTP error and the \canditatePatch is considered as invalid.

\paragraph{\regressionService}

Let us now consider the \regressionService.
In our simulation of bug Mayocat-231, the \regressionService performs regression testing on 80 synthetic requests (as per RQ2). However, it
does not reject patches based on this simulated workload.
This happens as the production traffic simulator is unable to create an input for this bug that requires regression testing (the bug report only provide us with crashing input). Note this is a limitation of our production traffic generator, not a conceptual limitation of \itzal. We note that designing generators of likely synthetic traffic is an unresearched area yet a very difficult problem.

\paragraph{Comparison against the Human Patch}

Among the 120 patches synthesized by \itzal which pass all oracles in this setup, none is syntactically equivalent to the patch written by the developer.
However, \autoref{lst:Mayocat-231:itzal_patch} shows an example of a semantically equivalent one, which has the same behavior as the human patch.
It replaces the null element (``carrier.getPerItem()'') by an existing variable ``BigDecimal.ZERO''found in the execution context. 

Note that, as shown by Long and Rinard  \cite{Long2016analysis}, it is common to have many equivalently correct yet syntactically different patches in the search space of a patch model.

\subsubsection{End-to-end Evaluation on BroadleafCommerce-1282}

We now study the case of bug Broadleaf Commerce-1282 (\url{https://github.com/BroadleafCommerce/BroadleafCommerce/issues/1282}), which is still in the domain of e-commerce. We focus on showing an important point that was not highlighted by the first case study: the fact that some aspects of patch optimality, in particular with respect to user experience, are not handled by standard validity oracles.

\begin{figure}[t]
\centering
\begin{lstlisting}[language=diff]
void populateEntityForm(...) {
    ...
    String idProperty = adminEntityService.getIdProperty(cmd);
    
    // null pointer exception here
    // because entity.findProperty(idProperty) is null when idProperty is not present in "entity"
    ef.setId(entity.findProperty(idProperty).getValue());
    ...
}
\end{lstlisting}
\caption{The failure point of bug BroadleafCommerce-1282.}
\label{lst:BroadleafCommerce-1282:failure_point}
\end{figure}

\begin{figure}[t]
\centering
\begin{lstlisting}[language=diff]
    adminInstance.setUsername(adminInstance.getEmailAddress());
    if (customerService.readCustomerByUsername(adminInstance.getUsername()) != null) {
-       Entity error = new Entity();
-       error.addValidationError("username", "nonUniqueUsernameError");
-       return error;
+       entity.addValidationError("emailAddress", "nonUniqueUsernameError");
+       return entity;
    }
\end{lstlisting}
\caption{The human patch for bug BroadleafCommerce-1282 (simplified version).}
\label{lst:BroadleafCommerce-1282:patch}
\end{figure}

\begin{figure}[t]
\centering
\begin{lstlisting}[language=diff]
@@ FormBuilderServiceImpl.java
@@ -717,2 +717,5 @@
    String idProperty = adminEntityService.getIdProperty(cmd);
+   if (entity.findProperty(idProperty) == null) {
+     return;
+   }
    ef.setId(entity.findProperty(idProperty).getValue());
\end{lstlisting}
\caption{The \itzal patch for bug BroadleafCommerce-1282.}
\label{lst:BroadleafCommerce-1282:bikini_patch}
\end{figure}

\paragraph{Description of the bug}

This bug is a null dereference that happens when the website administrator adds a customer with an email address that already exists in the database (i.e., the email address is already used by another customer).
When this failure occurs, the user interface displays a low level debugging stack trace. Contrary to bug Mayocat-231 that completely breaks the website, this bug has a lower severity.

\autoref{lst:BroadleafCommerce-1282:failure_point} shows the failure point (i.e., where the null pointer exception happens): When idProperty is ``emailAddress'', ``entity.findProperty(idProperty)'' returns null as no such property exists in the entity. Consequently, the call to ``getValue()'' results in a null pointer exception. 

\paragraph{\itzal patches}
\itzal generates 12 different compilable \canditatePatches with NPEFix patch model, the other patch model did not succeed to generate patch for this bug (Again, the complete list of \itzal patches is available at \cite{repo}). 
Among the 12 patches, 5 of them avoid the null pointer exception and do not produce any new bad behaviors that are detected by the \requestOracle.

Let us analyze one of them as shown in \autoref{lst:BroadleafCommerce-1282:bikini_patch}. This patch handles the failure by exiting the method when utility method ``findProperty'' does not find the required property.
With this patch, no dirty error message is displayed in the user interface and can be considered as a valid workaround to the problem.

\paragraph{Comparison against the Human Patch}

When comparing the \itzal patch against the human patch, the surprise is that they are in different methods. The human patch (shown in \autoref{lst:BroadleafCommerce-1282:patch}) is in method ``validateUniqueUsername'', and it essentially replaces the error identifier ``username'' by ``emailAddress''. Later on, at the failure point, the ``emailAddress'' property that is looked up is found and no exception is thrown.

From the viewpoint of the \requestOracle (the absence of exception in this case study), both patches handle the failure and both are correct. However, the human patch is conceptually better. While the \itzal patch silently skips the action to be done and gives no feedback to the user, the human patch transforms the exception into a clean and explicit warning about duplicate emails. This shows that there are cases where the absence of domain knowledge in the patch model and/or in the oracle results in sub-optimal patches. To overcome this problem, the developer always has the option to improve the patches shown in the \itzal dashboard before merging them in the code base of the application.

\answer{4}{This end-to-end experiment shows the feasibility of  deploying the novel program repair scheme \itzal on real applications. 
It also highlights that the main challenge of doing this kind of research in the laboratory is to have good workloads reflecting production traffic.}

\section{Related Work}
\label{sec:rw}

\subsection{Test Suite Based Patch Generation}
\label{subsect:rw-patch}

The literature on test suite based patch generation is growing very fast, and we here only present a brief overview of notable contributions.
GenProg \cite{le2012genprog} applies genetic programming to the AST of a buggy program and generates patches by adding, deleting, or replacing AST nodes. Prophet \cite{prophet} and Genesis \cite{long2017automatic} learn from existing successful human patches to improve the repair success rate. SemFix \cite{nguyen2013semfix} is a constraint based repair approach. It provides patches for assignments and conditions by combining symbolic execution and code synthesis. DirectFix \cite{mechtaev2015directfix} and Angelix \cite{angelix} from the same team provide further improvement over SemFix. Nopol \cite{nopol} is also a constraint based method, which focuses on repairing bugs in if-conditions and missing preconditions. There are also some template-based patch generation approaches. PAR \cite{Kim2013} uses 10 patch templates for common programming errors and Relifix \cite{reliflix} defines templates specifically for regression bugs. 

Main novelty: \itzal does patch synthesis directly in production, which none of those contributions address. \itzal does not require a failing test case or a strong regression test suite.

\subsection{Runtime Repair in Production}

Several automatic repair techniques handle failures in production, we now review the notable ones.

Rx \cite{qin2005rx} is a runtime repair system based on changing the environment upon failures. Rx employs checkpoint-and-rollback for re-executing the buggy code when failures happen. 
Assure \cite{sidiroglou2009assure} is a self-healing system also based on checkpointing.
For both of them, there is no patch generation strategy associated with the checkpoint and rollback mechanism.

Rinard et al. \cite{rinard2004enhancing} present a technique called ``failure oblivious computing'' to avoid illegal memory accesses by adding additional code to each memory operation during the compilation process.
Dobolyi and Weimer~\cite{dobolyi2008changing} present a technique to tolerate null dereferences by introducing hooks to inject manufactured values.
Long et al. \cite{long2014automatic} also explore this idea with the concept of ``recovery shepherding''.
Upon certain errors (null dereferences and divide by zero), recovery shepherding consists of returning manufactured values for failure oblivious computing.
Perkins et al. \cite{perkins2009automatically} propose ClearView, a system for automatically repairing errors in production by monitoring the system execution on low-level registers to learn invariants. 
Gu et al. \cite{gu2016automatic} present Ares, a runtime error recovery technique for Java exceptions using JavaPathFinder (JPF). 

Main novelty: these techniques do not directly produce source code patches that are communicated to developers. On the contrary, \itzal provides a fully automated bridge between production and source code patches for developers.

\subsection{Shadow Traffic}
The concept of shadow traffic is related with the execution of multiple versions of the same software in parallel, called in the literature ``multi-version execution'' \cite{hosek2013safe}, ``parallel execution'' \cite{Trachsel2010ParallelExecution} or simply ``dual execution'' \cite{Kwon2016LDX} when only two versions run. However, none of these work aims to generate patches. Using shadow traffic for repair has only recently been suggested in a short vision paper \cite{durieux2016production}, which remains at the concept level.

\emph{For security.~}
Kwon et al. \cite{Kwon2016LDX} do dual execution for detecting information leaks and attacks.
Salamat et al. \cite{salamat09} do multi-version execution also for the sake of security, and they implement the monitor entirely in userspace.
In both cases, patch generation is not considered.
The idea of shadow traffic is closely related to the idea of shadow executions introduced by Capizzi et al. \cite{Capizzi2008shadowexecution}. However, the goals are completely different. While they aim at isolating all private information, our goal is to perform patch search. 

\emph{For performance.~}
Trachsel and Gross \cite{Trachsel2010ParallelExecution} perform parallel execution to speed up programs. The instances that are run in parallel are different implementations of the same algorithm or different binary versions compiled with different optimization options. Compared to \itzal, no actionable feedback is given to the developer.

\emph{For reliability.~}
Hosek and Cadar \cite{hosek2013safe} do multi-version execution over versions and switch between versions when a bug is detected.
This technique can handle failures because some bugs may disappear between different versions. 
It is not clear how this technique can be used for patch generation.

Main novelty: to our knowledge, \itzal is the first approach to use production traffic for validating automatically synthesized patches.

\section{Conclusion}
\label{sec:conclusion}

In this paper, we have presented \itzal, an approach for synthesizing patches live for production failures. 
This novel and disrupting scheme for program repair is based on the conjunction of embedding the patch search in production, together with validating the absence of regressions based on the whole, diverse, production usages and values.
We have evaluated our novel technique based on \generationNbBug failures.

This new line of research in automatic software repair calls for future work.
First, there is a need to research on how to efficiently synchronize an application and its shadows (the mirror applications fed with the shadow traffic). Second, we envision a feedback loop with developers as follows.
When a developer discards or modifies a generated patch, this information should be given back to \itzal, then 
the \patchService would automatically refine the patch model, the \regressionService would automatically synthesize better execution comparison oracles, and finally, the patch prioritization done in the dashboard would be the result of a machine learning approach.

\balance

\bibliographystyle{IEEEtran}
\bibliography{references}

\balance
\end{document}